\documentclass{PoS}

\usepackage{graphicx}
\usepackage{overpic}

\usepackage{amsmath}
\usepackage{amssymb}
\usepackage{picins}
\usepackage{mathbbol}
\usepackage{slashed}

\newcommand{\units}[1]{\ensuremath{\,\mathrm{#1}}}
\newcommand{\bra}[1]{\left\langle #1 \right|}
\newcommand{\ket}[1]{\left| #1 \right\rangle}

\newcommand{\mRe}{\mathrm{Re}\,}
\newcommand{\mIm}{\mathrm{Im}\,}

\newcommand{\crossout}[1]{\setbox0\hbox{#1} \hbox to \wd0{\rlap{\rule[.6ex]{\wd0}{.5pt}}\box0}} 
\newcommand\topalignbox[1]{\leavevmode\vtop{\vskip-0ex\hbox{#1}}}
\newcommand\mtopalignbox[1]{\leavevmode\vtop{\vskip-12pt\hbox{#1}}}

\newcommand{\elll}{\ell}

\newcommand{\mqquad}{\hspace{3mm}}
\newcommand{\mquad}{\hspace{2mm}}

\newcommand{\mybox}[1]{{#1}}

\title{Transverse Momentum Distributions of Quarks from the Lattice using Extended Gauge Links}

\ShortTitle{Transverse Momentum Distributions using Extended Gauge Links}

\author{%
	\speaker{Bernhard~U.~Musch}$^a$,
	Philipp~H\"agler$^a$,
	Andreas~Sch\"afer$^b$,
	Meinulf~G\"ockeler$^b$,
	Dru~B.~Renner$^c$,
	John~W.~Negele$^d$,
	LHPC (Lattice Hadron Physics Collaboration)\\
	\llap{$^a$}	Institut f\"ur Theoretische Physik T39, Physik-Department, Technische Universit\"{a}t M\"{u}nchen, \\
	James-Franck-Stra{\ss}e, D-85747 Garching, Germany\\
	\llap{$^b$}Institut f\"ur Theoretische Physik, Universit\"at Regensburg, \\
	D-93040 Regensburg, Germany\\
	\llap{$^c$}Department of Physics, University of Arizona, \\
	1118 E 4th Street, Tucson, AZ 85721, USA\\
	\llap{$^d$}Center for Theoretical Physics, Massachusetts Institute of Technology, \\
		Cambridge, MA02139, USA\\
	E-mail: \email{bmusch@ph.tum.de}%
	}

\newcommand{\mskt}{\ensuremath{\langle \vec k_T^2 \rangle}}

\abstract{We present preliminary numerical studies in Lattice QCD related to the intrinsic transverse momentum distribution of partons in the nucleon. We employ non-local operators, consisting of spatially separated quark creation and annihilation operators connected by a straight Wilson line. A clear signal is already obtained from a small number of configurations at a pion mass $m_\pi\approx 600$ MeV. As an example, we demonstrate that we can obtain the first $x$-moment of the transverse momentum dependent parton distribution function $f_1^{n=1}(\vec k_T)$ from our data. Our results, which are not 
renormalized, show a Gaussian-like distribution. The root mean squared transverse momentum is $\sqrt{\mskt} \approx 560\units{MeV}$ for a Gaussian fit, close to phenomenological values.}

\FullConference{The XXV International Symposium on Lattice Field Theory\\
                July 30 - August 4 2007\\
                Regensburg, Germany}

\begin{document}

% \todo{check abbreviation TMPDF; Ji: TMDPD}

\section{Introduction}

% A nucleon at relativistic speed in a scattering experiment can be described in terms of the parton picture: When a quark (or gluon) is hit by the target, it behaves like a ``parton'', i.e. it adopts a fraction $x$ of the nucleon four-momentum. In addition, the momentum of the parton has a transverse component $\vec{k}_T$, perpendicular to the velocity of the nucleon.The distribution of quarks with respect to $x$ and/or $\vec{k}_T$ is encoded in transverse momentum dependent parton distribution functions (TMDPDFs) $f_1(x,\vec{k}_T)$, $h_{1}(x,\vec{k}_T)$, $\ldots$ \cite{Mulders:1995dh}. Experimentally, information about the intrinsic transverse momentum of partons becomes relevant in semi-inclusive deep inelastic scattering (SIDIS) and Drell-Yan processes.

The Bjorken-$x$-dependence of parton distribution functions (PDFs) has been investigated by several lattice collaborations in recent years. In these studies, the lattice operators probing the nucleon always represent a \emph{local} continuum operator. Spatial separations between the quark annihilation and creation operator on the lattice only appear in the context of discretized covariant derivatives.

% We would now like to extend the repertoire of lattice observables, resolving the dependence on the intrinsic parton momentum $\vec{k}_T$ as well. 
In this study, we explore the possibility to examine the dependence on the intrinsic parton momentum $\vec{k}_T$ as well. 
For the purpose of calculating transverse momentum dependent parton distribution functions (TMDPDFs), we investigate \emph{non-local} operators, constructed from quark and antiquark fields which are spatially separated. Gauge invariance is ensured by introduction of a straight gauge link (Wilson line) connecting the quark fields.

\section{TMDPDFs in SIDIS experiments}

\begin{figure}[bp]
	\begin{center}
	\begin{minipage}{0.48\textwidth}
		\mybox{a)\\ \topalignbox{\includegraphics[width=\textwidth]{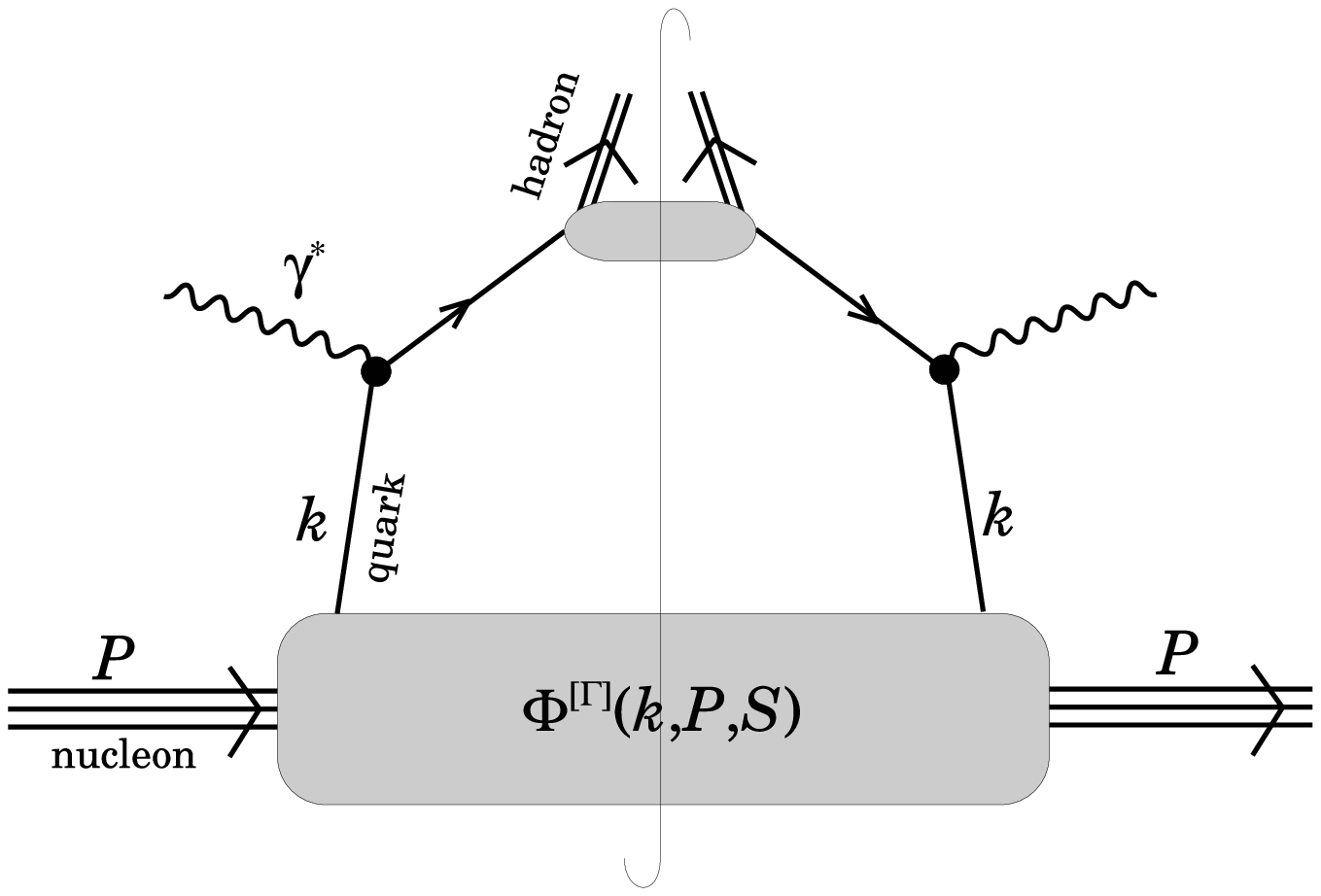}}}
		\end{minipage} \hfill
	\begin{minipage}{0.48\textwidth} \begin{flushright}
		\mybox{b)\mquad\topalignbox{\includegraphics*[width=0.80\textwidth]{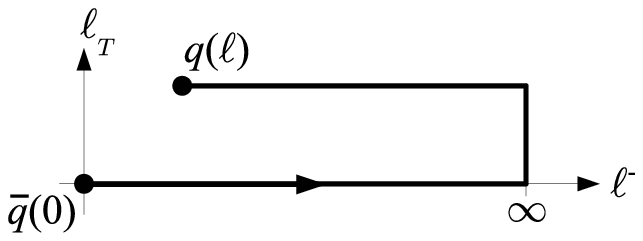}}}\vspace{2mm}\\%
		\mybox{c)\mquad\topalignbox{\includegraphics*[width=0.80\textwidth]{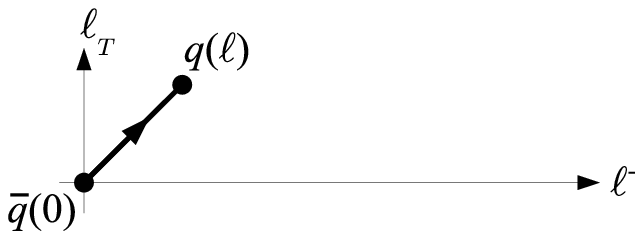}}}%
		\end{flushright} \end{minipage} 
	\end{center}
	\caption{a)\mquad Factorized tree level diagram of semi-inclusive deep inelastic scattering (SIDIS),\mqquad b)\mquad Gauge link to infinity and back as in SIDIS,\mqquad c)\mquad Straight gauge link \label{fig-softblobs}}
\end{figure}

In experimental processes like semi-inclusive deep inelastic scattering (SIDIS), one factorizes dominant diagrams into hard, perturbative processes and soft, non-perturbative parts \cite{Collins:1983ju,Mulders:1995dh,Ji:2004wu}, represented as shaded areas in fig.~\ref{fig-softblobs}\,a). The lower soft part can be parameterized in terms of nucleon TMDPDFs and corresponds to a correlator
\begin{align}
	\displaystyle \Phi^{[\Gamma]}(k,P,S)\ & \equiv\ \int \frac{d^4 \elll}{2 (2 \pi)^4}\ e^{i k\cdot \elll}\ \bra{P,S}\ \mathcal{O}^{\Gamma}(\elll)\  \ket{P,S}\ , \label{eq-fundcorr} \\ \text{where} \hspace{9mm}
	\mathcal{O}^{\Gamma}(\elll) & \equiv \overline{q}(0)\, \Gamma\, \mathcal{U}_{[0,\elll]}\, q(\elll) \label{eq-contop}	
\end{align}
%where $\mathcal{O}^{\Gamma}(\elll)  \equiv \overline{q}(0)\, \Gamma\, \mathcal{U}_{[0,\elll]}\, q(\elll)$. 
Here $\ket{P,S}$ represents a nucleon state of momentum $P$ and spin $S$, $\Gamma$ is a Dirac matrix and $k$ is the quark momentum. $\mathcal{U}_{[0,\elll]}$ is a gauge link connecting the quark operators. 
% \begin{equation}
% 	\mathcal{U}(0 \rightarrow \elll)\ \equiv\ \mathcal{P} \exp \left( -i g \int_0^{\elll} d\xi^\mu A_\mu(\xi) \right)
% 	\end{equation}
For SIDIS, the gauge link runs to infinity and back, as illustrated in fig.~\ref{fig-softblobs}\,b), see, e.g., ref.~\cite{Boer:2003cm}. 

\pagebreak
\section{First approach to TMDPDFs from the lattice}

\piccaption{Step-like link path for $\vec l = (6,3,0)$. \label{fig-steplike}}
\parpic[l][l]{
	\mybox{\includegraphics[width=.25\textwidth]{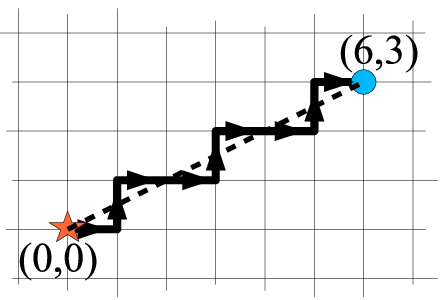}}% 
	}
The idea of this study is to extract TMDPDFs from matrix elements $\bra{P,S}\ \mathcal{O}^{\Gamma}(\elll)\  \ket{P,S}$ calculated directly on the lattice.
%We have now begun to develop a method to extract the distribution associated to the transverse momentum $\vec{k}_T$.
So far we restrict ourselves to straight gauge link operators: in the continuum limit, the gauge link follows a straight path connecting quark annihilation and creation operators, see fig.~\ref{fig-softblobs}\,c). The matrix elements thus calculated do not correspond exactly to those relevant for SIDIS, but the resulting TMDPDFs
have an interesting probabilistic interpretation \cite{Ralston:1979ys}. 
%\noindent On the lattice, the operator $\mathcal{O}^{\Gamma}(\elll)$ is implemented as
% \begin{equation}
%	\mathcal{O}^{\Gamma}_\text{lat}(\vec \elll, \tau) \equiv \frac{1}{2}
%	\sum_{\vec{z}}\ 
%	\overline{q}(\vec z,\tau)\ \Gamma\
%	\mathcal{U}\!(\vec{z} \rightarrow \vec{z}+\vec{\elll})\
%	q(\vec z+\vec \elll,\tau) 	
%	\end{equation}
%where now the gauge link $\mathcal{U}\!(\vec{z} \rightarrow \vec{z}+\vec{\elll})$ is a product of link variables. 
On the lattice, the gauge link $\mathcal{U}_{[0,\elll]}$ in the operator  $\mathcal{O}^{\Gamma}_\text{lat}(\elll)$ is a product of link variables.
Since we work in Euclidean space-time, %on the lattice, 
we can only evaluate spatial quark separations $\elll = (0, \vec{\elll})$. Therefore we restrict our lattice operators to a single slice $\tau$ in Euclidean time.
For quark separation $\vec{l}$ which do not lie on the $x$, $y$ or $z$ axis, we can approximate a straight connection with a step-like link path, see fig.~\ref{fig-steplike} (inset). \par

\begin{figure}[bp]
	\mybox{\begin{minipage}{0.48\textwidth}
		a)\mquad
		\mtopalignbox{\includegraphics*[width=0.9\textwidth]{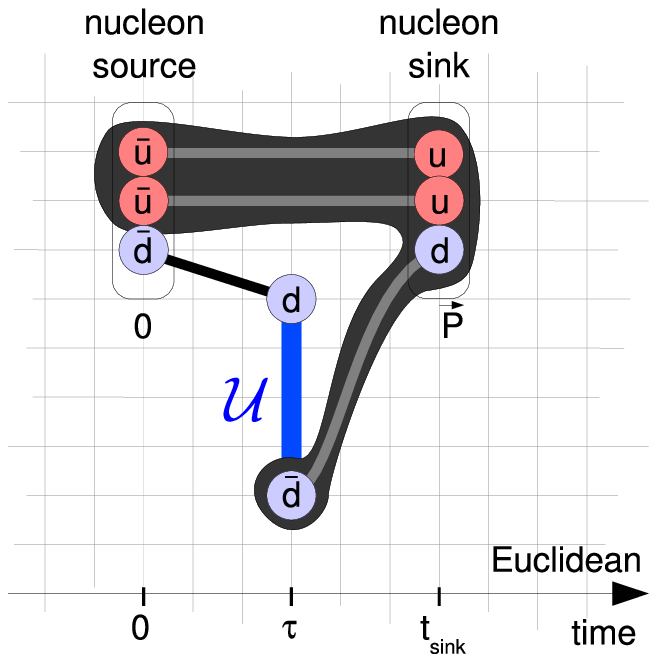}}
		\end{minipage}}\hfill
	\mybox{\begin{minipage}{0.48\textwidth}
		b)\mquad
		\mtopalignbox{\includegraphics*[width=0.9\textwidth]{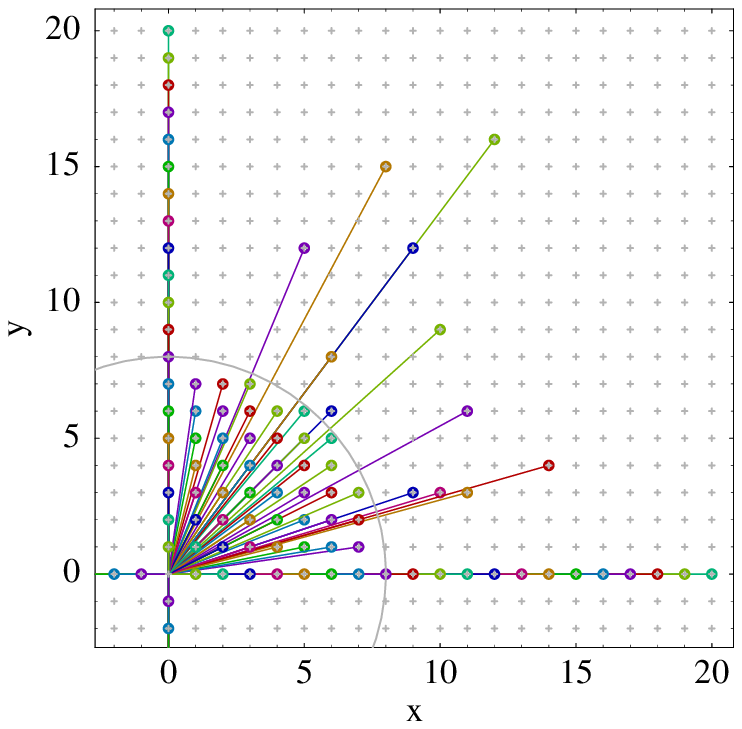}}
		\end{minipage}}
	\caption{a)\mquad Evaluation of the three-point function in the numerator of eq.~(\protect\ref{eq-ratiodef}) on the lattice (schematic), here for an operator $\mathcal{O}^{\Gamma}_\text{lat}$ with d-quarks. Only one of the two possible connected contractions of quark fields is shown. All-to-all propagators are avoided by combining three of the quark propagators into a sequential propagator (dark area).\mqquad b)\mquad Overview of quark separations $\vec{\elll}$ in the x-y-plane used in this investigation. We have calculated all quark separations which lie on the x- or y-axis, up to a length of 20 lattice units. In the first quadrant, we have included all quark separations up to a length of 8 lattice units (inner grey circle) and a selection of longer ones. \mqquad  \label{fig-method}}
\end{figure}

% The matrix element can be written as $ \bra{P,S}\ \mathcal{O}^{\Gamma}(\elll)\  \ket{P,S} = \overline{U}(P,S)\ \tilde{\mathcal{M}}_{\Gamma}(\elll,P)\ U(P,S)$, and $\tilde{\mathcal{M}}_{\Gamma}(\elll,P)$ can be parameterized in terms of Lorentz-invariant structures.
To extract a signal for nucleon matrix elements of the operators in eq.\,({\ref{eq-contop}}), we form ratios of nucleon three-point and two-point functions 
\begin{equation}
	R_\Gamma(\tau;\vec{P},\vec{\elll}) = \frac{\Gamma^{\text{3pt}}_{\alpha \beta}\ \langle\  B_\beta(t_{\text{sink}},\vec{P})\ \mathcal{O}^{\Gamma}_\text{lat}(\vec \elll; \tau)\ \overline{B}_\alpha(0,\vec{P})\ \rangle }
	{ \Gamma^{\text{2pt}}_{\alpha \beta}\ \langle\  B_\beta(t,\vec{P})\ \overline{B}_\alpha(0,\vec{P})\ \rangle } ,
	\label{eq-ratiodef}
\end{equation}
% \begin{align}
% 	C^{\text{2pt}}(t,\vec P) &= \Gamma^{\text{2pt}}_{\alpha \beta}\ 
% 	\langle\  B_\beta(t,\vec{P})\ \overline{B}_\alpha(t_{\text{src}},\vec{P})\ \rangle \\
% 	C^{\text{3pt}}_\mathcal{O}(\tau,\vec P) & = \Gamma^{\text{3pt}}_{\alpha \beta}\ 
% 	\langle\  B_\beta(t_{\text{snk}},\vec{P})\ \mathcal{O}\ \overline{B}_\alpha(t_{\text{src}},\vec{P})\ \rangle 
% 	\label{eq-twothreept}
% 	\end{align}
where  $\Gamma^{\text{2pt}}$ and $\Gamma^{\text{3pt}}$ are suitable nucleon spin projection matrices\footnote{LHPC uses $\Gamma^{\text{2pt}}=\frac{1}{2}(\Eins + \gamma_4)$ and $\Gamma^{\text{3pt}}=\frac{1}{2}(\Eins + \gamma_4)(\Eins + i \gamma_5 \gamma_3)$}, and where $B(t,\vec{P})$ is a nucleon interpolating operator, composed of $u$ and $d$ quark operators. Nucleon source and sink are placed at times $t_{\text{source}} = 0$ and $t_{\text{sink}}$, respectively. The transfer matrix formalism reveals that at a sufficiently large distance from source and sink ($0 \ll \tau \ll t_\text{sink}$), the ratio $R_\Gamma(\tau;\vec{P},\vec{\elll})$ becomes $\tau$-independent. This plateau value is directly related to the value of the matrix element $\bra{P,S}\ \mathcal{O}^{\Gamma}_\text{lat}(\elll)\  \ket{P,S}$. 

% If operator, source and sink are sufficiently far apart, the transfer matrix formalism predicts 
% \begin{equation}
% 	R_{\Gamma}(\tau;\vec{P},\vec{\elll}) \mathop{\approx}^{\phantom{M}0 \ll \tau \ll t_\text{sink} \phantom{M}} \frac{1}{2 E(P)} \frac{\mathrm{Tr}\  (\slashed{P}+m_N)\, \Gamma^\text{3pt}\, (\slashed{P}+m_N) \, \tilde{\mathcal{M}}_{\Gamma}^\text{lat}(\elll,P)}{ \mathrm{Tr}\ (\slashed{P}+m_N)\, \Gamma^\text{2pt} }
% 	\end{equation}
% where $E(P)$ is the energy and $m_N$ the mass of the nucleon.
% Thus we obtain $\tilde{\mathcal{M}}_{\Gamma}^\text{lat}(\elll,P)$ from the plateau value of the ratio.

For the evaluation of eq.~(\ref{eq-ratiodef}), we apply the standard technique as described, e.g., in ref.~\cite{Dolgov:2002zm}, based on products of propagators and sequential propagators as illustrated in fig.~\ref{fig-method}\,a). In the following, we present results for isovector operators ($q=u-d$), where disconnected contributions are absent.

\section{Test setup}

\newcommand{\numconfigs}{84}

In our first numerical test calculations, we use \numconfigs\ MILC gauge configurations from the NERSC archive \cite{Orginos:1998ue,Orginos:1999cr}. The configurations were produced with the AsqTad improved staggered quark action with 2+1 flavors. The lattice dimensions are $L^3 \times T = 20^3 \times 64$, with a lattice spacing $a \approx 0.124 \units{fm}$. The quark masses are $a m_{u,d} = 0.030$, and $a m_s = 0.050$. The gauge configurations have been HYP smeared and bisected in the temporal direction, and we have selected only the time slices $0\ldots31$. We are using unsmeared propagators and sink-smeared sequential propagators previously calculated by the LHPC group for these chopped gauge configurations (see, e.g. \cite{Hagler:2007xi}). The propagators have been calculated using domain wall fermions, with the quark mass tuned such that the pion mass $m_\pi \approx 596 \units{MeV}$ is approximately equal to the Goldstone pion mass for the staggered sea quark action. The source-sink separation is $t_\text{sink} - t_\text{source} = 10$. LHPC has calculated sequential propagators for two nucleon momenta $\vec P=(0,0,0)$ and $\vec P=(-1,0,0)$. The latter corresponds to a momentum of $500\units{MeV}$ in physical units.

We have explored a number of link paths in all directions. Fig.~\ref{fig-method}\,b) illustrates our selection of quark separations $\vec{\elll}$ in the $x$-$y$-plane. For our test runs, we have chosen two Dirac structures in the operator: the vector case $\Gamma=\gamma_4$ and the axial vector case $\Gamma=\gamma_3 \gamma_5$.

\begin{figure}[bp]
	\begin{center}
	\mybox{\begin{minipage}{0.47\textwidth}
		a)\\ \mtopalignbox{\includegraphics*[width=\textwidth]{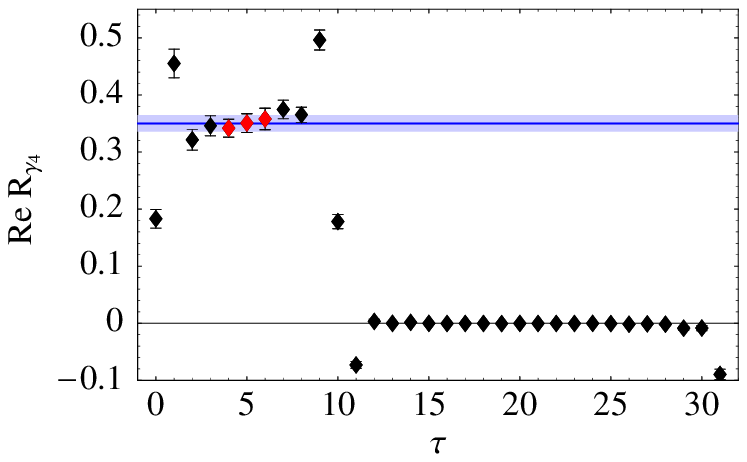}} 
		\end{minipage}} \hfill
	\mybox{\begin{minipage}{0.47\textwidth}
		b)\\ \mtopalignbox{\includegraphics*[width=\textwidth]{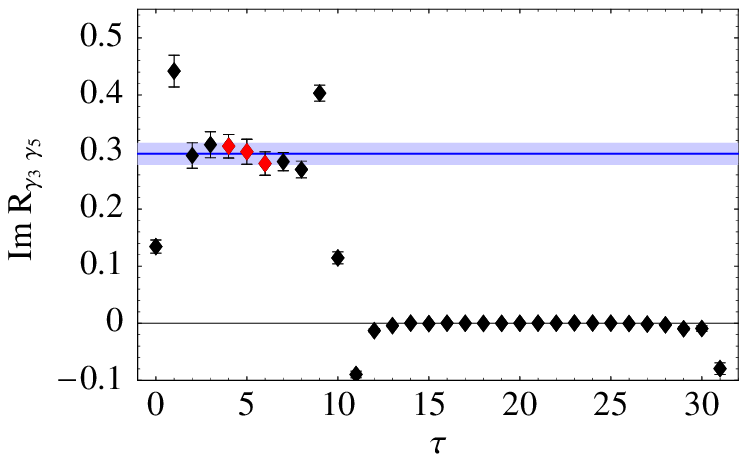}}
		\end{minipage}} 
	\caption{Sample plateau plots: $R_{\Gamma} (\tau; \vec P,\vec \elll)$ is plotted versus $\tau$. The horizontal line and the error band indicate the plateau value and its error, extracted from the points (marked red), at $\tau = 4,5$ and $6$.\mqquad a)\mquad Real part of $R_\Gamma (\tau; \vec P,\vec \elll)$ for $\Gamma = \gamma_4$, nucleon momentum $\vec P=(0,0,0)$, and a link path five units long in $x$ direction,\mquad i.e. $|\vec \elll|=5$.\mqquad b)~\mquad Imaginary part of $R_{\Gamma} (\tau, \vec P,\vec \elll)$ for  $\Gamma = \gamma_3 \gamma_5$, \mquad nucleon momentum $\vec P=(0,0,0)$,\mquad and the link path shown in fig.~\protect \ref{fig-steplike}, i.e. $|\vec \elll|=6.7$. \label{fig-plateau}}
	\end{center}
\end{figure}

\section{Preliminary Results}

% In the following, statistical errors are estimated applying the Jacknife-method. Fits are repeated for each Jacknife sample, in order to respect correlations among the data. The fits are standard least-squares fits, with fit weights adjusted to the Jacknife errors of the individual data points. \par

In fig. \ref{fig-plateau} we show two sample plots of the ratio $R_{\Gamma}(\tau; \vec P,\vec \elll)$ versus $\tau$. We obtain  clean plateaus. In order to extract the plateau value $R_{\Gamma}(\vec P,\vec \elll)$, we average over time slices at $\tau = 4,5$ and $6$. \par

\begin{figure}[bp]
	\mybox{\begin{minipage}{0.47\textwidth}
		a) \\ \mtopalignbox{\includegraphics[width=\textwidth]{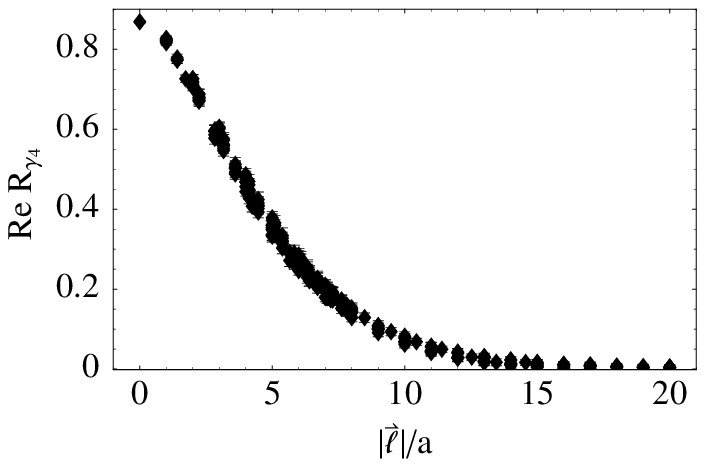}}
		\end{minipage}} \hfill
	\mybox{\begin{minipage}{0.47\textwidth}
		b) \\ \mtopalignbox{\includegraphics*[width=\textwidth]{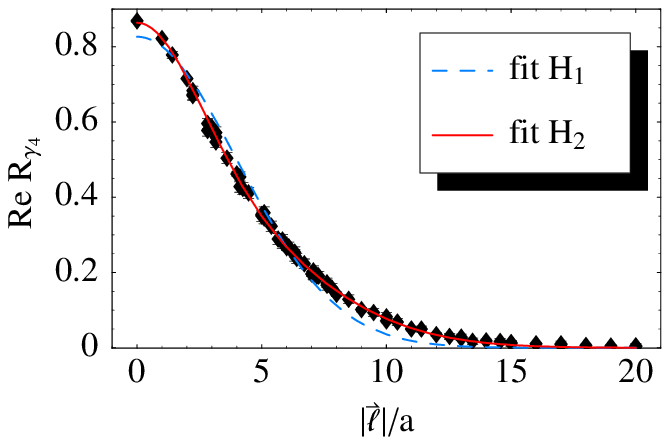}}
		\end{minipage}} \hfill
	\caption{ a)\mquad Results for $\Gamma=\gamma_4$, $\vec P=(0,0,0)$. We plot $\mRe R_{\gamma_4}(\vec P,\vec \elll)$ for all link paths versus the separation $|\vec \elll|$ of quark creation and annihilation operator. \mqquad b)\mquad $\mRe R_{\gamma_4}(\vec P = 0,\vec \elll)$ versus $|\vec \elll|$ for link paths in the x-y-plane. Results for link paths which transform into one another under rotation or reflection have been averaged. Dashed turquoise curve: fit to the data with a single Gaussian function $H_1(|\vec \elll|)$, see eq.\,(\protect \ref{eq-gauss}). Solid red curve: fit with the superposition of two Gaussian functions $H_2(|\vec \elll|)$. The parameters determined from the fits are listed in tables~\protect \ref{tab-fitresultssingle} and \protect\ref{tab-fitresultsdouble}.
	\label{fig-ratios} }
\end{figure}

Fig. \ref{fig-ratios}\,a) shows all results $R_{\Gamma}(\vec P,\vec \elll)$ for $\Gamma=\gamma_4$ and $\vec P=(0,0,0)$ for the 263 evaluated link paths. 
The signal is quite good, even for longer quark separations.
We find that the correlator primarily depends on the separation $|\vec \elll|$ between quark annihilation and creation operator.

\begin{figure}[bp]
	\begin{center}
	\mybox{\begin{minipage}{0.47\textwidth}
		a) \\ \mtopalignbox{\includegraphics[width=\textwidth]{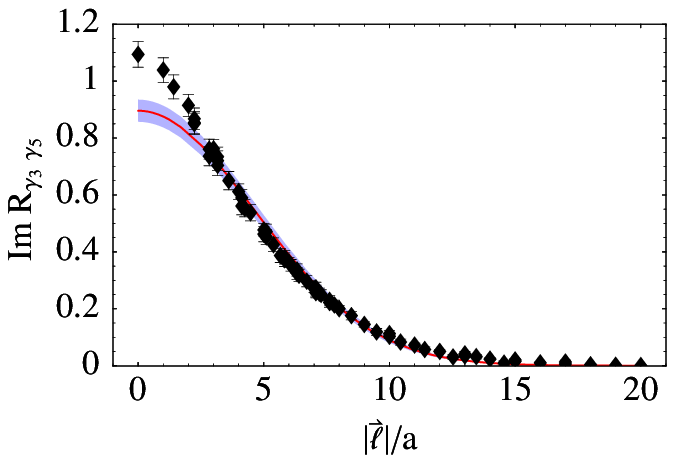}}
		\end{minipage}} \hfill
	\mybox{\begin{minipage}{0.47\textwidth}
		b) \\ \mtopalignbox{\includegraphics[width=\textwidth]{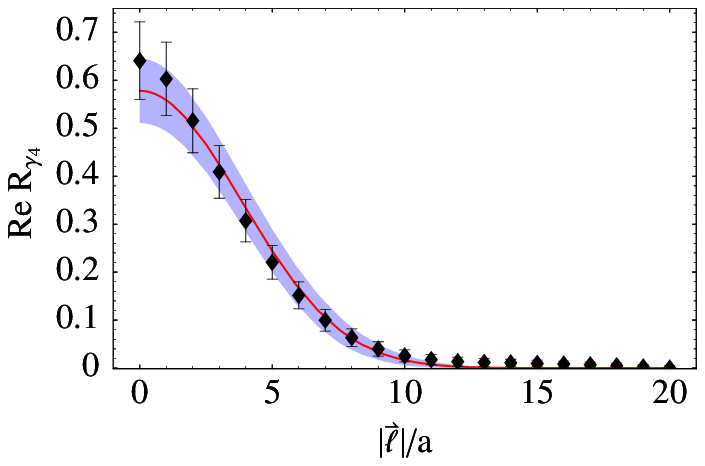}}
		\end{minipage}} 
	\caption{a)\mquad $\mIm R_{\Gamma}$ versus $|\vec \elll|$ for $\Gamma=\gamma_3 \gamma_5$ (axial vector), momentum $\vec P=(0,0,0)$ and link paths in the x-y-plane. Results for link paths which transform into one another under rotation or reflection have been averaged.  \mqquad b)\mquad  $\frac{1}{2} \mRe \{R_\Gamma(\vec\elll) + R_\Gamma(-\vec\elll) \} $ for $\Gamma=\gamma_4$,\mquad$\vec \elll$ on the positive $x$-axis, and non-zero nucleon momentum $\vec P=(-1,0,0)$.\mquad In both plots, the data have been fitted with a single Gaussian function $H_1(|\vec \elll|)$, see eq.\,(\protect \ref{eq-gauss}). Parameters determined from the fits are listed in table~\protect \ref{tab-fitresultssingle}. \label{fig-moreratios}}
	\end{center}
\end{figure}

In figures \ref{fig-ratios}\,b) (vector case, unpolarized) and \ref{fig-moreratios}\,a) (axial vector case, polarized) we select only link paths lying in the $x$-$y$-plane. Furthermore, we identify groups of link paths which transform into one another under rotation or reflection in the $x$-$y$-plane, and take the group average. 
In fig. \ref{fig-moreratios}\,b) we show an example at non-zero nucleon momentum. For quark separations $\vec \elll$ on the positive $x$-axis, nucleon momentum $\vec P = (-1,0,0)$ and $\Gamma=\gamma_4$, we plot $\frac{1}{2} \mRe \left\{R_\Gamma(\vec\elll) + R_\Gamma(-\vec\elll) \right\} $. Here we refrain from averaging over the whole $x$-$y$-plane, because the parameterization indicates that the value is not invariant with respect to the link direction.

We have tested the following fit functions to parameterize the $|\vec \elll|$ dependence:
\begin{equation}
	H_1(|\vec \elll|) := C_1 \exp\left( - \frac{|\vec \elll|^2}{\sigma_1^2} \right)\ , \hspace{8mm}
	H_2(|\vec \elll|) := C_1 \exp\left( - \frac{|\vec \elll|^2}{\sigma_1^2} \right) 
                  + C_2 \exp\left( - \frac{|\vec \elll|^2}{\sigma_2^2} \right)\ . \label{eq-gauss}
\end{equation}
The resulting fit parameters are listed in tables~\ref{tab-fitresultssingle} and \ref{tab-fitresultsdouble}. We observe that the data can be well described by the sum of Gaussians in $H_2$.

\begin{table}[bp]

	\begin{tabular}{l||c|c||c|c|c}
		fig. & $\Gamma$ & $\vec P$ & $C$ & $\sigma$ & $2/\sigma$ \\ \hline
		\ref{fig-ratios}b)\,dashed & $\gamma_4$ & $(0,0,0)$ & 
		$0.826 \pm 0.005$ & $(5.64 \pm 0.12) a = 0.70\units{fm}$ & $(563 \pm 12) \units{MeV} $ \\
		\ref{fig-moreratios}\,a) & $\gamma_3 \gamma_5$ & $(0,0,0)$ & 
		$0.90 \pm 0.04$  & $(6.58 \pm 0.12) a = 0.82\units{fm}$ & $(484 \pm 9)\units{MeV}$ \\
		\ref{fig-moreratios}\,b) & $\gamma_4$ & $(-1,0,0)$ & 
		$0.58 \pm 0.07$ & $(5.4 \pm 0.5) a = 0.67\units{fm}$ & $(589 \pm 46)\units{MeV}$
		\end{tabular} \\

\caption{Fit parameters determined from the single Gaussian fits with $H_1$ shown in fig. \protect \ref{fig-ratios}\,b) and \protect \ref{fig-moreratios}.}
\label{tab-fitresultssingle}
\end{table}

\begin{table}[bp]

	\begin{tabular}{l||c|c||c|c|c||c|c|c}
		fig. & $\Gamma$ & $\vec P$ & $C_1$ & $\sigma_1$ & $2/\sigma_1$ & $C_2$ & $\sigma_2$ & $2/\sigma_2$ \\ \hline
		\ref{fig-ratios}b)\,solid & $\gamma_4$ & $(0,0,0)$ & 
		$0.49 $ & $7.3 a$ & $(433 \pm 15) \units{MeV} $ &
		$0.37 $ & $3.4 a$ & $(945 \pm 41) \units{MeV} $
		\end{tabular} \\

\caption{Fit parameters determined from the double Gaussian fit with $H_2$ shown in fig. \protect \ref{fig-ratios}\,b) }
\label{tab-fitresultsdouble}
\end{table}

\section{A first glimpse of TMDPDFs from the lattice}

\begin{figure}[bp]
	\begin{centering}
	\includegraphics[width=0.47\textwidth]{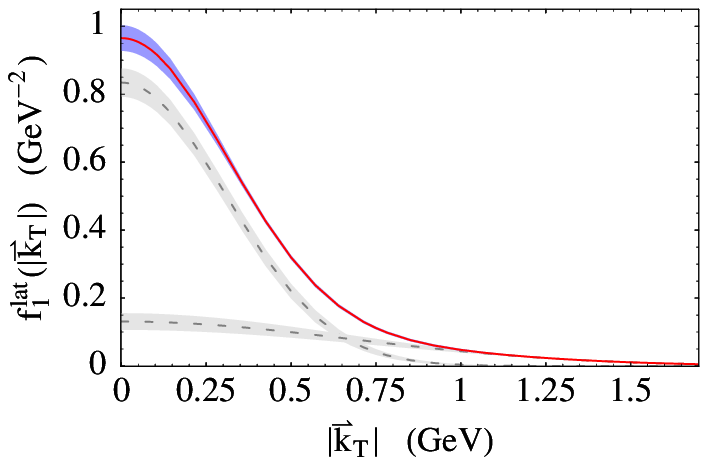}\par
	\end{centering}
	\caption{Upper curve: $f_{1,\text{lat}}^{n=1}(\vec k_T)$, calculated from a Fourier transform of $R_{\gamma_4}(\vec P = 0,|\vec \elll|)$. The upper curve is the sum of the two dashed curves, which show the Fourier transforms of the two Gaussian components in fit function $H_2(|\vec \elll|)$. The results are not
	% yet
	renormalized.\label{fig-ft}}
	\end{figure}

We can use the parameters of the fits in fig.~\ref{fig-ratios}\,b) to calculate the first $x$-moment ($n=1$) of the intrinsic transverse momentum dependence of the parton density $f_1$:
\begin{equation}
	f_1^{n=1}(\vec k_T) \equiv \int_{-1}^1 dx \int_{-\infty}^{\infty} dk^- \left( \int \frac{d^4 \elll}{2 (2 \pi)^4}\ e^{i k\cdot \elll}\ \bra{P,S}\ \overline{q}(0)\, \gamma^+\, \mathcal{U}_{[0,\elll]}\, q(\elll)  \ket{P,S} \right)\ . 
	\end{equation}
It turns out that
\begin{equation}
	f^{n=1}_{1,\text{lat}}(\vec k_T) = \int \frac{d^2 \vec \elll_T}{(2 \pi)^2} e^{-i\,\vec k_T \cdot \vec \ell_T} R_{\gamma_4} ( \vec P = 0,|\vec \ell_T|)\ ,
\end{equation}
where ``lat'' indicates that the operator has not been renormalized. %, yet.
For the single Gaussian fit (function $H_1$), we find a root mean square transverse momentum of $\sqrt{\langle {\vec k_T}^2 \rangle} = 2/\sigma = (563 \pm 12)\units{MeV}$, see table~\ref{tab-fitresultssingle}. This is very well compatible with a value of $\approx 500\units{MeV}$ used in recent phenomenological investigations of HERMES data on SIDIS \cite{Anselmino:2005nn}, based on the factorized Ansatz $f_1(x, \vec k_T) = f_1(x) \exp[ - \vec k_T^2 / \mskt] / [\pi \mskt] $.
%Model calculations typically assume that the intrinsic momentum distribution of quarks in the nucleon is Gaussian. As an example, the authors of ref.~\cite{Anselmino:2005nn} make the ansatz $f_1(x, \vec k_T) = f_1(x) \exp[ - \vec k_T^2 / \mskt] / [\pi \mskt] $, and find that a value $\sqrt{\mskt} = 500 \units{MeV}$ matches best with HERMES data.
Note however, that such a comparison with phenomenological results has to be taken with due caution, since the effect of renormalization of the non-local operators $\mathcal{O}^{\Gamma}_\text{lat}$ could, in principle, affect the $\vec 
\elll$ dependence of $R_{\Gamma}$. 
The result for the double Gaussian fit (function $H_2$) is shown in fig.~\ref{fig-ft} and table~\ref{tab-fitresultsdouble}. 
We obtain a root mean square of the transverse momentum of $\sqrt{\langle {\vec k_T}^2 \rangle} = (702 \pm 12)\units{MeV}$.

\section{Conclusions and outlook}

We have calculated nucleon matrix elements $\bra{P,S}\ q(0)\, \Gamma\, \mathcal{U}_{[0,\elll]}\, q(\vec{\elll})\  \ket{P,S}$ with a finite separation $\elll$ of the quark operators. It turns out that the dependence on $\vec{\elll}$ is approximately Gaussian in the channels we explored. We have used our data to obtain a first, preliminary result on transverse momentum dependent parton distribution functions (TMDPDFs). The root mean square transverse momentum $\sqrt{ \langle{\vec k_T}^2 \rangle}$ of our unrenormalized result for $f_1^{n=1}(\vec k_T)$ is $(563 \pm 12)\units{MeV}$ for a single Gaussian fit, a value which is compatible with phenomenological results. It will be interesting to study the correlations with respect to the quark separation $\vec \elll$ and the nucleon momentum $\vec P$. We also plan to investigate whether there is a lattice analogy to link paths extending to infinity and back. This would enable us to calculate the TMDPDFs directly relevant to phenomenology.

\acknowledgments

Thanks are due to Vladimir~Braun for helpful discussions
%. We are grateful 
and to the members of the LHPC collaboration for providing propagators and technical expertise. B.~M. and Ph.~H. acknowledge support by the DFG Emmy Noether-program and A.~S. by BMBF. This work was supported in part by funds provided by the U.S. Department of Energy under grant DE-FG02-94ER40818.

\end{document}